\documentclass[conference]{IEEEtran}
\ifCLASSINFOpdf
\else
\fi
\usepackage{amsfonts}
\usepackage{mathrsfs}
\usepackage[ruled,linesnumbered]{algorithm2e}

\usepackage{amsmath}
\usepackage{xcolor}

\usepackage{graphicx}
\usepackage{epstopdf}
\ifCLASSOPTIONcompsoc
    \usepackage[caption=false, font=normalsize, labelfont=sf, textfont=sf]{subfig}
\else
\usepackage[caption=false, font=footnotesize]{subfig}
\fi

\SetKwInput{KwInit}{Initialize}
\SetKwInput{KwMain}{Main Loop}

\begin{document}
\title{Dynamic Detection of False Data Injection Attack in Smart Grid using Deep Learning}

\author{\IEEEauthorblockN{Xiangyu Niu}
\IEEEauthorblockA{Department of Electrical Enginnering \\and Computer Science\\
University of Tennessee, Knoxville\\
Knoxville, Tennessee 37996\\
Email: xniu@vols.utk.edu}
\and
\IEEEauthorblockN{Jiangnan Li}
\IEEEauthorblockA{Department of Electrical Enginnering \\and Computer Science\\
	University of Tennessee, Knoxville\\
	Knoxville, Tennessee 37996\\
	Email: jli103@vols.utk.edu}
\and
\IEEEauthorblockN{Jinyuan Sun}
\IEEEauthorblockA{Department of Electrical Enginnering \\and Computer Science\\
	University of Tennessee, Knoxville\\
	Knoxville, Tennessee 37996\\
	Email: jysun@utk.edu}}


%Telephone: (800) 555--1212, Fax: (888) 555--1212}


\maketitle
\begin{abstract}
Modern advances in sensor, computing, and communication technologies enable various smart grid applications. The heavy dependence on communication technology has highlighted the vulnerability of the electricity grid to false data injection (FDI) attacks that can bypass bad data detection mechanisms.
Existing mitigation in the power system either focus on redundant measurements or protect a set of basic measurements. These methods make specific assumptions about FDI attacks, which are often restrictive and inadequate to deal with modern cyber threats.
In the proposed approach, a deep learning based framework is used to detect injected data measurement. Our time-series anomaly detector adopts a Convolutional Neural Network (CNN) and a Long Short Term Memory (LSTM) network. To effectively estimate system variables, our approach observes both data measurements and network level features to jointly learn system states. 
The proposed system is tested on IEEE 39-bus system. Experimental analysis shows that the deep learning algorithm can identify anomalies which cannot be detected by traditional state estimation bad data detection. 
\end{abstract}

\section{Introduction}
The future smart grid is designed to operate more reliable, economical and efficient in an environment of increasing power demand. This goal, however, is achieved by incorporating with a tremendous increase of data communications which lead to great opportunities for a various of cyber attacks. Thus, ensuring cyber security of the Smart Grid is a critical priority. Although a large number of countermeasures have been published, such as communication standards (e.g. IEC 61850-90-5 \cite{iec61850}), regulation laws (e.g. Colorado Regulations (CCR) 723-3), cryptographic implementations (e.g. secure channel \cite{fouda2011lightweight}), and official guidelines (e.g. NISTIR 7628 Guidelines \cite{grid2010introduction}), current smart grid still remains vulnerable to cyber attacks.

% Data are measured by end meters such as (PMU and AMI) and transmitted through the network and processed by a state estimator to filter the measurement noise and to detect gross errors. The results of state estimation are then used by applications at the control center, for purposes such as contingency analysis, optimal power flow, economic dispatch, etc.. 

To prevent cyber attacks, legacy grid relies on traditional security scheme (e.g., firewall and general intrusion detection system). Intrusion detection systems (IDS) are able to generate alarms for potential intrusions by consistently monitoring network traffic or system logs. Although there are a number of studies on general IDS in network security community, limited effort has been made specifically to smart grid. At the same time, the risk of attacks targeting data availability and integrity in the power networks is indeed real. A notable work is by Liu et al. \cite{liu2011false}, which proposed a type of attacks, called false data injection (FDI) attacks, against the state estimation in the power grid. In such attacks, the attackers aim to bypass existing bad data detection system and pose damage on the operation of power system by intentional changing the estimated state of the grid systems. Therefore, there is an urgent need of effective smart grid specific intrusion detection systems.

To address the above issues, two schemes have been widely studied to detect FDI attacks \cite{bobba2010detecting, liang2017review}: One way is to strategically protect a number of secure basic measurements. Kim et al. \cite{kim2011strategic} propose a greedy algorithm to select a subset of base measurements and the placement of secure phasor measurement units. Bi et al. \cite{bi2014graphical} characterize the problem into a graphical defending mechanism to select the minimum number of meter measurements which cannot be compromised. The other way of defending FDI attack is to verify each state variables independently. Liu et al. \cite{liu2014detecting} formulate a low rank matrix separation problem to identify attacks and propose two optimization methods to solve the problem. Ashok et al. \cite{ashok2016online} present an online detection algorithm that utilizes statistical information and predictions of the state variables to detect measurement anomalies.

Recently, machine learning algorithms have been broadly adopted to the smart grid literature for monitoring and preventing cyber attack of power systems. Ozay et al. \cite{ozay2016machine} generate Gaussian distributed attacks and use both supervised and semi-supervised machine learning methods to classify attacks. 
Similarly, Esmalifalak et al. \cite{esmalifalak2014detecting} devise a distributed support vector machines based model for labeled data and a statistical anomaly detector for unsupervised learning cases.
He et al. \cite{he2017real} employs Conditional Deep Belief Network (CDBN) to efficiently reveal the high-dimensional temporal behavior features of the unobservable FDI attacks.
However, existing works mainly focus on finding bad measurement at certain state, no prior studies have been conducted over the dynamic behavior of FDI attack. Besides, detecting FDI attacks is considered as supervised binary classification problem in \cite{esmalifalak2014detecting, he2017real} which are incapable of detecting dynamically evolving cyber threats and changing system configuration.

Recent breakthrough in GPU computing provide the foundation for neural network to go "deep". In this paper, we develop an anomaly detection framework based on neural network to enable the construction of a smart grid specific IDS.  More specifically, a recurrent neural network with LSTM \cite{hochreiter1997long} cell is deployed to capture the dynamic behavior of power system and a convolutional neural network \cite{krizhevsky2012imagenet}  is adopt to balance between two input sources. An attack is alerted when residual between the observed and the estimated measurements is greater than a given threshold. 

Moreover, attackers with sophistic domain knowledge may continually manipulate the power grid state estimation without being detected causing extensive damages. As such, we want to bridge the gap between network anomaly detector and FDI attacks detection mechanism. Unlike other works which separate two detectors, our framework combines both network traffic characteristics and time-series data measurements with help of convolution neural network to equalize between two inputs. With the help of the proposed neural network structure, our anomaly detector demonstrates highly accurate detection performance.

We organize the rest of this section as follows: Section~\ref{sec:background} introduces the background of FDI attack and neural network. Section~\ref{sec:detection} presents our combined detection system along with the static and dynamic method to detect FDI attack in Section~\ref{sec:static} and Section~\ref{sec:dynamic}. Section~\ref{sec:eva} presents the case study on IEEE 10-machine 39-bus power system. Finally, we conclude our work in Section~\ref{sec:con}.

\section{Background} \label{sec:background}
\subsection{False Data Injection Attack}
In a power system, the state is represented by bus voltage magnitudes $V \in \mathcal{R}^n$ and angles $\theta \in ([-\pi, \pi])^n$, where $n$ is the number of buses. Let $z = [z_1, z_2, . . . , z_m] ^ T \in \mathcal{R} ^ m $ be the measurement vector, $x = [x_1, x_2, . . . , x_n] ^ T \in \mathcal{R}^n$ be the state vector, and $e = [e_1, e_2, . . . , e_m]^T \in \mathcal{R}^m$ denote the measurement error vector. We describe the AC measurement model as follows:

\begin{equation}
z = h(x) + e
\end{equation}

In analyzing the impact of data attack on state estimation, we adopt the DC model obtained by linearizing the AC model where the relationship between these $m$ meter measurements and $n$ state variables can be characterized by an $m \times n$ matrix $\mathbf{H}$. In general, the matrix $\mathbf{H}$ of a power system is a constant matrix determined by the topology and line impedances of the system.

\begin{equation}
z = \mathbf{H} x + e
\end{equation}

Typically, a weighted least squares estimation is used to obtain the state estimate as $\hat{x}= \min_x \frac{1}{2}(z - \mathbf{H} x)^T\mathbf{R}^{-1}(z - \mathbf{H} x) = (\mathbf{H}^T\mathbf{R}^{-1} \mathbf{H})^{-1}\mathbf{H}^T\mathbf{R}^{-1}z$ where $\mathbf{R}$ is the covariance matrix.

Let $z_a$ represent the vector of observed measurements that may contain malicious data. $z_a$ can be represented as $z_a = z + a$ where $a = (a_1, ..., a_m)^T$ is the malicious data added to the original measurements. Let $x_a$ denote the estimates of $x$ using the malicious measurements $z_a$. Then $x_a$ can be represented as $\hat{x} + c$, where $c$ is a non-zero vector representing the impact on the estimate from the malicious injection and $\hat{x}$ is the estimate using the original measurements. In this paper, for target FDI attackers, we assume the attacker has enough inside information to constructing $x_a$ while random FDI attackers only have partial information.

\subsection{Convolutional Neural Network}
\begin{figure}[!t]
	\centering
	\includegraphics[width=0.95\linewidth]{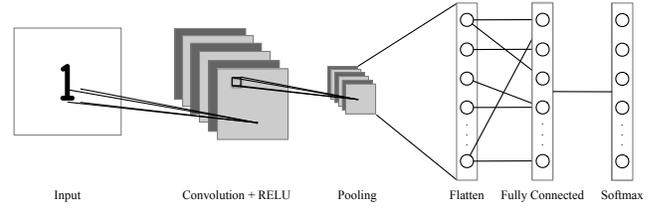}
	\caption{An example of architecture for image classification with convolutional neural network.}
	\label{fig:cnn}
\end{figure}

Convolutional Neural Networks (CNNs or ConvNets) are a category of Neural Networks that have been successful in processing image and video signal such as identifying objects in real time video and style transfer for images as visualized in Figure \ref{fig:cnn}.
% Formally, a one-hidden-layer MLP is a function:
% \begin{equation}
% f(x) = G(b^{(1,t)} + W^{(1,t)}x)
% \end{equation}
% where $G$ are the activation functions; $W^{(t)}$ and $b^{(t)}$ denote the bias vectors and the weight matrices. The output is then obtained as: $o(x) = G(b^{(2,t)} + W^{(2,t)} h(x))$

CNNs describe the most classic form of neural network where multiple nodes are arranged in layers such that information only follows from input to output. We use three main types of layers to build a CNN architecture: Convolutional Layer, Pooling Layer, and Fully-Connected Layer.

% \begin{itemize}
% \item Convolutional layer computes the output of neurons that are connected to local regions in the input, each computing a dot product between their weights and followed by a non-linear function where we often use rectified linear unit (ReLU) $f(z) = max(0, z)$.
% \item Pooling layer reduces the dimensionality of each feature map but retains the most important information. Spatial pooling can be of different types: Max, Average, Sum etc. 
% \item The fully connected layer is a one-hidden-layer MLP that uses a softmax activation function (multi-classification task) or sigmoid activation function (binary classification task) in the output layer.
% \end{itemize}

In this way, CNN transforms the original input layer by layer from the original tensor to the final output which can be class score. In particular, each convolutional layer and fully connected layers perform transformation that is a function of both the parameters (weights and biases) and activations in the input volume. The parameters in the CNN will be trained with gradient descent optimization algorithm to minimize the loss between the outputs that the CNN computes and the labels of training dataset. 

% \begin{figure}[h]
% 	\centering
% 	\includegraphics[width=0.55\linewidth]{fig/mlp.eps}
% 	\caption{A multi-layer perceptron (MLP) with one hidden layer.}
% 	\label{fig:mlp}
% \end{figure}

\subsection{Recurrent Neural Network}
Recurrent neural networks, or RNNs \cite{rumelhart1986learning}, are a family of neural networks for processing sequential data. Contrasting from convolutional network which is specialized for processing high dimensional tensors such as image, a recurrent neural network is a neural network that is specialized for processing time-series values $x^{(1)}, ... , x^{(t)}$. RNNs are ideal for long sequences without sequence-based specialization. 

% \begin{figure}[h]
% 	\centering
% 	\includegraphics[width=0.95\linewidth]{fig/rnn.eps}
% 	\caption{A recurrent neural network with no outputs.}
% 	\label{fig:rnn}
% \end{figure}

Recurrent neural networks use the following equation to define the values of  hidden units. 

\begin{equation} \label{equ:rnn}
h^{(t)}= f(h^{(t-1)}, x^{(t)}; \theta)
\end{equation}
where $h$ represent the state and $x^{(t)}$ refers the time-series input at time $t$. %This expression can be represented by a directed computational graph as illustrated in Figure \ref{fig:rnn}.

% When the recurrent network is trained to perform a task that requires predicting the future from the past, the network typically learns to use $h^{(t)}$ as a kind of loss $y$ summary of the task-relevant aspects of the past sequence of inputs up to $t$. This summary is in general necessarily loss $y$, since it maps an arbitrary length sequence $(x^{(t)}, x^{(t−1)}, ..., x^{(1)})$ to a fixed length vector $h(t)$. Depending on the training criterion, this summary might selectively keep some aspects of the past sequence with more precision than other aspects. For example, if the RNN issued in statistical language modeling, typically to predict the next word given previous words, storing all the information in the input sequence up to time $t$ may not be necessary; storing only enough information to predict the rest of the sentence is sufficient.

\begin{figure}[!t]
	\centering
	\includegraphics[width=0.85\linewidth]{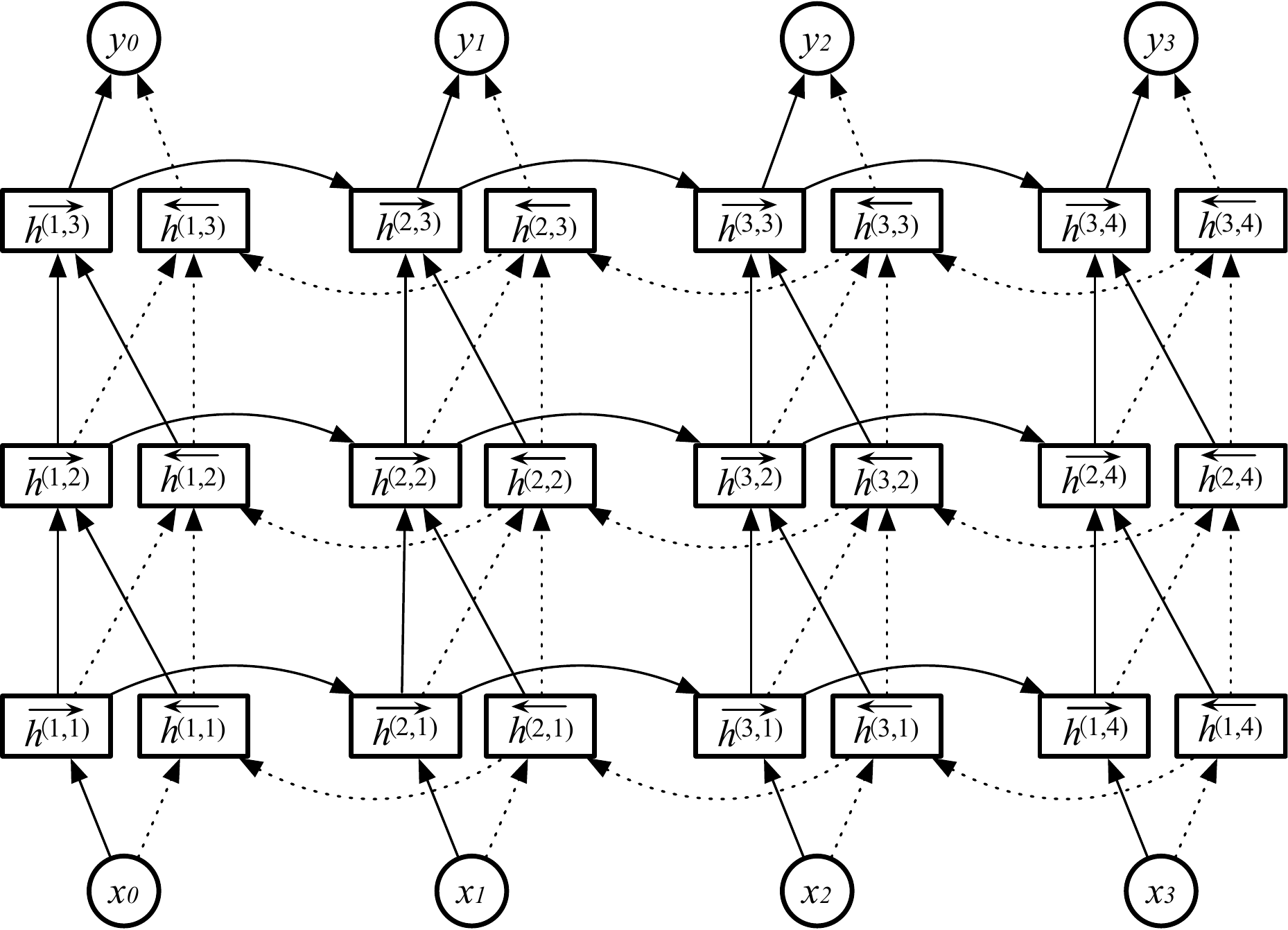}
	\caption{Computation of a typical 3 layers bidirectional recurrent neural network.}
	\label{fig:biRNN}
\end{figure}

Schuster et al. \cite{schuster1997bidirectional} shows a bi-directional deep neural network. At each time-step $t$, bi-directional RNN maintains two hidden units, one for the forward propagation and another for the backward propagation. The final result, $y^t$, is generated through combining the score results produced by both hidden units. Figure \ref{fig:biRNN} shows the bi-directional network architecture, and \eqref{equ:birnn} show the formulation of a single bidirectional RNN hidden layer.
\begin{align}\label{equ:birnn}
\overrightarrow{h}{(t)} &= f(h^{(t-1)}, x^{(t)}; \overrightarrow{\theta})\\
\overleftarrow{h}{(t)} &= f(h^{(t+1)}, x^{(t)};\overleftarrow{\theta})\label{equ:birnn2}
\end{align}

\subsubsection{Long Short-Time Memory}
Currently, the most commonly implemented RNNs fall into the class of long short-time memory (LSTM) neural networks \cite{hochreiter1997long}. Different from vanilla RNN with single gate, LSTM exhibits notable performance gain for preserving long time dependencies while also keeping short time memories. Each LSTM cell involve three gates to which are input gate $i$, output gate $o$, the forget gate $f$. The information flow of LSTM cell is as follows:
\begin{align}
f_{t}&=\sigma _{g}(W_{f}x_{t}+U_{f}h_{t-1}+b_{f})\\
i_{t}&=\sigma _{g}(W_{i}x_{t}+U_{i}h_{t-1}+b_{i})\\
o_{t}&=\sigma _{g}(W_{o}x_{t}+U_{o}h_{t-1}+b_{o})\\
c_{t}&=f_{t}\circ c_{t-1}+i_{t}\circ \sigma _{c}(W_{c}x_{t}+U_{c}h_{t-1}+b_{c})\\
h_{t}&=o_{t}\circ \sigma _{h}(c_{t})
\end{align}
where $\sigma_g(\cdot)$ and $\sigma_c(\cdot)$ represent the sigmoid and tangent function respectively, and $\circ$ denotes the element-wise
product. Here, $c$ and $h$ stand for the cell state vector and hidden unit vector. 

% \begin{figure}[h]
% 	\centering
% 	\includegraphics[width=0.7\linewidth]{fig/lstm.eps}
% 	\caption{The block diagram of the LSTM recurrent network.}
% 	\label{fig:lstm}
% \end{figure}

% \subsubsection{Bidirectional RNN}
% RNNs look into the past words to predict the next word in the sequence. It is possible to make predictions based on future words by having the RNN model read through the corpus backwards. 

\section{Detecting Data Level False Data Injection Attack} \label{sec:detection}

\begin{figure}[!t]
    \centering
    \subfloat[Limited power FDI attacks ]{\label{fig:scen1}\includegraphics[width=0.45\linewidth]{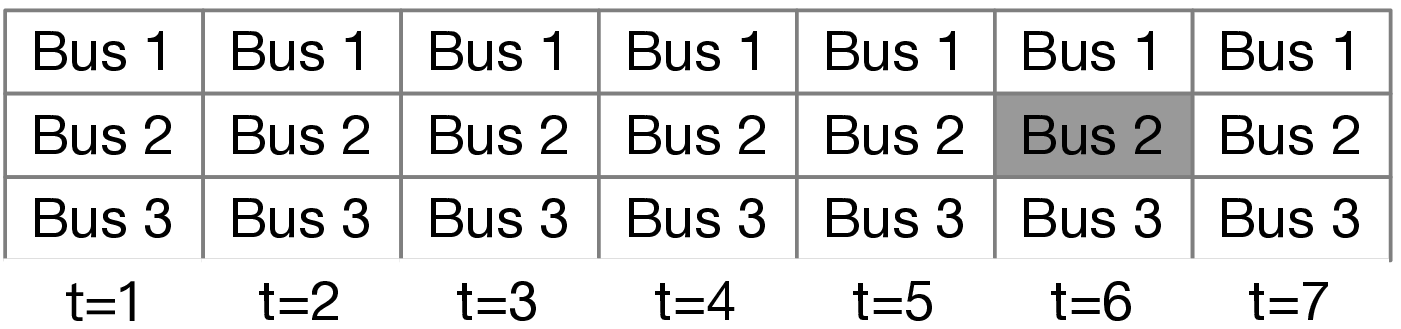}} \hfill
    \subfloat[Unlimited power FDI attacks]{\label{fig:scen2}\includegraphics[width=0.45\linewidth]{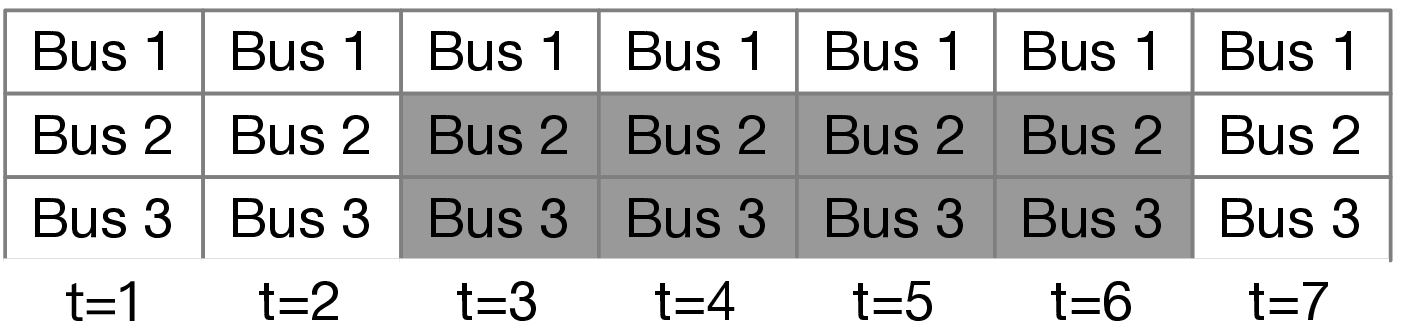}} \hfill
    \caption{Different scenario for dynamic FDI attacks.}
    \label{fig:dynamic_scen}
\end{figure}

Various research on static FDI attack detection method has been published. A common assumption is a threat model where the attackers have knowledge of the power system topology; however, can only inject a limited number of bad data points which is shown in Figure. \ref{fig:scen1}. In this threat model, FDI attack can be mitigated if a proportion of the comprised substation is below a certain threshold. Moreover, data measurements are often redundant for estimating the actual state. This threat model is widely adopted in existing works. Nonetheless, we stress this threat mode by: 1) removing the limitation of the number of measurement data that are corrupted; and 2) assuming the attackers have basic understanding of the aforementioned static detection mechanism in \eqref{equ:sta_detect}.

Figure \ref{fig:scen2} shows the dynamic FDI attack that is focused in this work. The attack starts at $t = 3$ and the measurements of both bus 2 and 3 have been compromised. Static method will fail in this scenario, for the reason that two thirds of the measurements have been modified from $t=3$ to $t=6$. A sophisticate attacker can deliberately generate a false event based on a real event and inject it to the power grid. As the result, it is unlikely to detect this attack only based on static method which can cost catastrophe results if control center makes false actions. 

\subsection{The Combined Attack Detection Method}\label{sec:combined}
\begin{figure}[!t]
	\centering
	\includegraphics[width=0.95\linewidth]{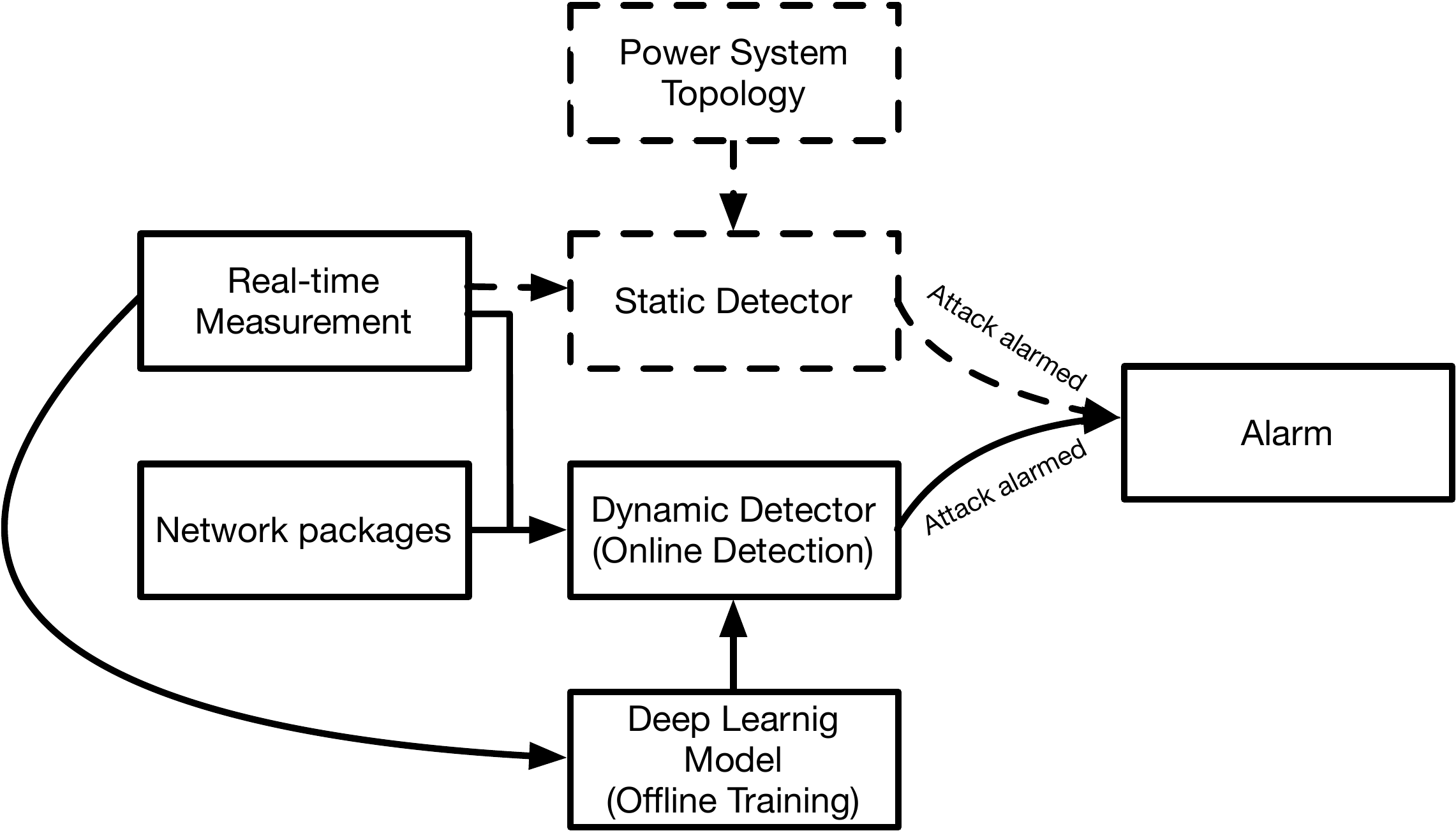}
	\caption{The overview of our proposed deep learning based FDI attacks detection system.}
	\label{fig:system}
\end{figure}

In this section, we provide an overview of our proposed system for detecting FDI attacks in Figure \ref{fig:system}. Our proposed detection mechanism mainly consists of a static detector and a deep learning based detection scheme. The static detector can be an State Estimator (SE) or any aforementioned FDI attack detector \cite{ashok2016online, bi2014graphical, esmalifalak2014detecting, he2017real, liu2014detecting, ozay2016machine, kim2011strategic} which is built independently beyond our dynamic detector.
As mentioned in the previous section, the dynamic detector takes two input sources. 
While the data level features are explicit, the network packages are captured by tcpdump and each network packet includes header and data payload, with unique features which defined in NSL-KDD dataset \cite{tavallaee2009detailed}. The NSL-KDD dataset has 41 features which are categorized into three types of features: basic, content based and traffic-based features. It should also be mentioned that some features are generated based on a fixed window (default is 2 second) which will remain consistent within the window. 

Our dynamic detector is employed to recognize the high-level time-series features of the FDI attacks. To achieve this goal, our time-series method consists of two essential mechanisms: offline training and online detection. The offline training is trained based on historical measurement and can be potentially facilitated by outsourcing to public machine learning cloud services. Unlike other methods which are designed under the assumption that the physical status of the power system does not change overtimes, our system will collect real-time measurement data to support offline training and the prediction model will update after retrain is completed.

\subsection{State Estimator Method} \label{sec:static}
%A naive detection scheme is to calculate the distance between the measurement $z$ and $\mathbf{H} x$. A typical solution is to use $\mathit{l}_2$-norm and a threshold $\tau$ is carefully chose to check whether there exist bad measurements or not. 
%
%\begin{equation}\label{equ:sta_detect}
%\begin{split}
%\Arrowvert \mathbf{H} x + e \Arrowvert _2 \leq \tau \\
%\Arrowvert \mathbf{H} x + e \Arrowvert _2 > \tau
%\end{split}
%\end{equation}
%
%However, this method can be easy mitigated by choosing $a$ as a linear combination of the column vectors of $H$. Which means, if the attacker is able to use $Hc$ as the attack vector $a$, $z_a$ can pass the detection as long as regular measurement $z$ can pass the detection. The proof can be characterized as follows:
%
%\begin{equation}
%\begin{split}
%\Arrowvert z_a - \mathbf{H} x\Arrowvert _2 &= \Arrowvert z + a - \mathbf{H} (x +c)\Arrowvert _2\\
%&=\Arrowvert z + a - \mathbf{H} x - \mathbf{H} c)\Arrowvert _2\\
%&=\Arrowvert z - \mathbf{H} x\Arrowvert _2 + \Arrowvert a - \mathbf{H} c\Arrowvert _2 \leq \tau
%\end{split}
%\end{equation}

In bad data injection attack, one method is to use Chi-squares test. Once bad data are detected, they need to be eliminated or corrected, in order to obtain the correct states.
There are two important hypothesis that are the largest normalized residual (LNR) and the $J(\hat{x})$ performance index:

\begin{equation}\label{equ:sta_detect}
J(\hat{x}) < \tau
\end{equation}

where $J(\hat{x})$ follows a chi-square distribution and $\tau$ is a preset threshold. The threshold can be obtained from the $\chi^2$ distribution. If $J(\hat{x}) > \tau$, bad data will be suspected. For DC model, $diag(\sigma ^{-2}_i , 0) = I$, the traditional bad data detection approaches often reduce to $l2$-norm of the measurement residual \cite{liu2011false}:

\begin{equation}
\Arrowvert z - \mathbf{H} \hat{x}\Arrowvert _2 < \tau
\end{equation}

\subsection{Dynamic Detection Method} \label{sec:dynamic}

In \cite{hao2016likelihood}, the authors formulate the bus voltage magnitudes, angles and states of measuring devices together as system states in Markov Decision Process (MDPs). In our method, we extend to a recursive model where the decision not only depends previous one state but previous $n$ states where the loss is as follows:

\begin{align} \label{equ:loss}
\eta &= L(\phi (s_t), f( \phi(s_{t-1}, ......, s_{t-n-1}), \theta), \tau)
\end{align}
where $\phi, \theta$ are parameters need to be turned and $\tau$ is the threshold that is needed to decide whether the attack has been started.

\begin{figure}[!th]
	\centering
	\includegraphics[width=0.85\linewidth]{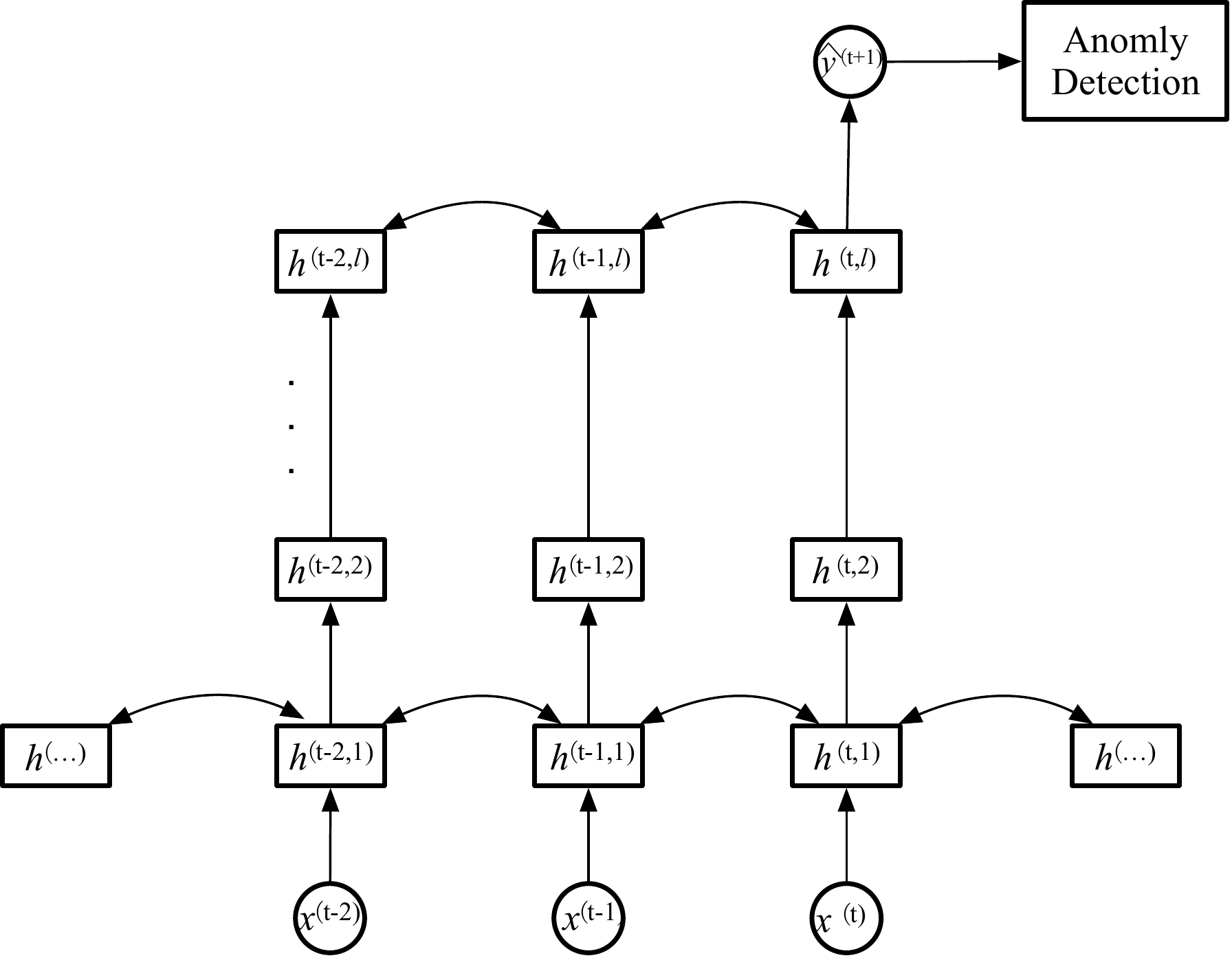}
	\caption{The time-serious dynamic detection method based on RNN.}
	\label{fig:dynamic_detection}
\end{figure}

Figure \ref{fig:dynamic_detection} shows the structure of our stacked dynamic detection model. Specifically, the input of the model are the time-serious power system data, the features will be passed to several LSTM layers to learn high dimensional temporal features. Previous works \cite{feng2017multi, he2017real} characterize FDI attacks as a binary classification problem which looks promising in the experimental setting, since the datasets to be tested can be manually tuned for different scenarios. In real world implementations, power system data is highly unbalanced, thus, binary classification methods will inevitably have low recall even the overall accuracy is high.  However, for evaluating IDS, recall is often more important than accuracy since any cyber attack can cost catastrophe results.

In general, our dynamic anomaly detector takes the input of time-series ${...,x^{(t-1)},x^{(t)},...}$, learn their higher dimensional feature representations, and then use those features to predict the next data point $\hat{x} ^{(t)}$. Furthermore, the predicted data point can be used to classify if $x(t)$ is anomalous by checking the similarity between the actual data $x(t)$ and predicted data $\hat{x} ^{(t)}$. 

Having presented single source FDI attacks detection model, we now introduce a framework that combines FDI attacks detector with network intrusion detection system. This framework is dealing with a case when an IDS that relies on data measurement fails to detect the start of FDI attacks. Accordingly, if the fabricated injection data are derived from a legitimate measurement in our threat model, data level detectors may fail to determine if current network is intruded or not. In this case, to increase the overall performance of time-series anomaly detection model, a combined attack detection method is proposed in this paper. 

\begin{figure}[!t]
	\centering
	\includegraphics[width=0.95\linewidth]{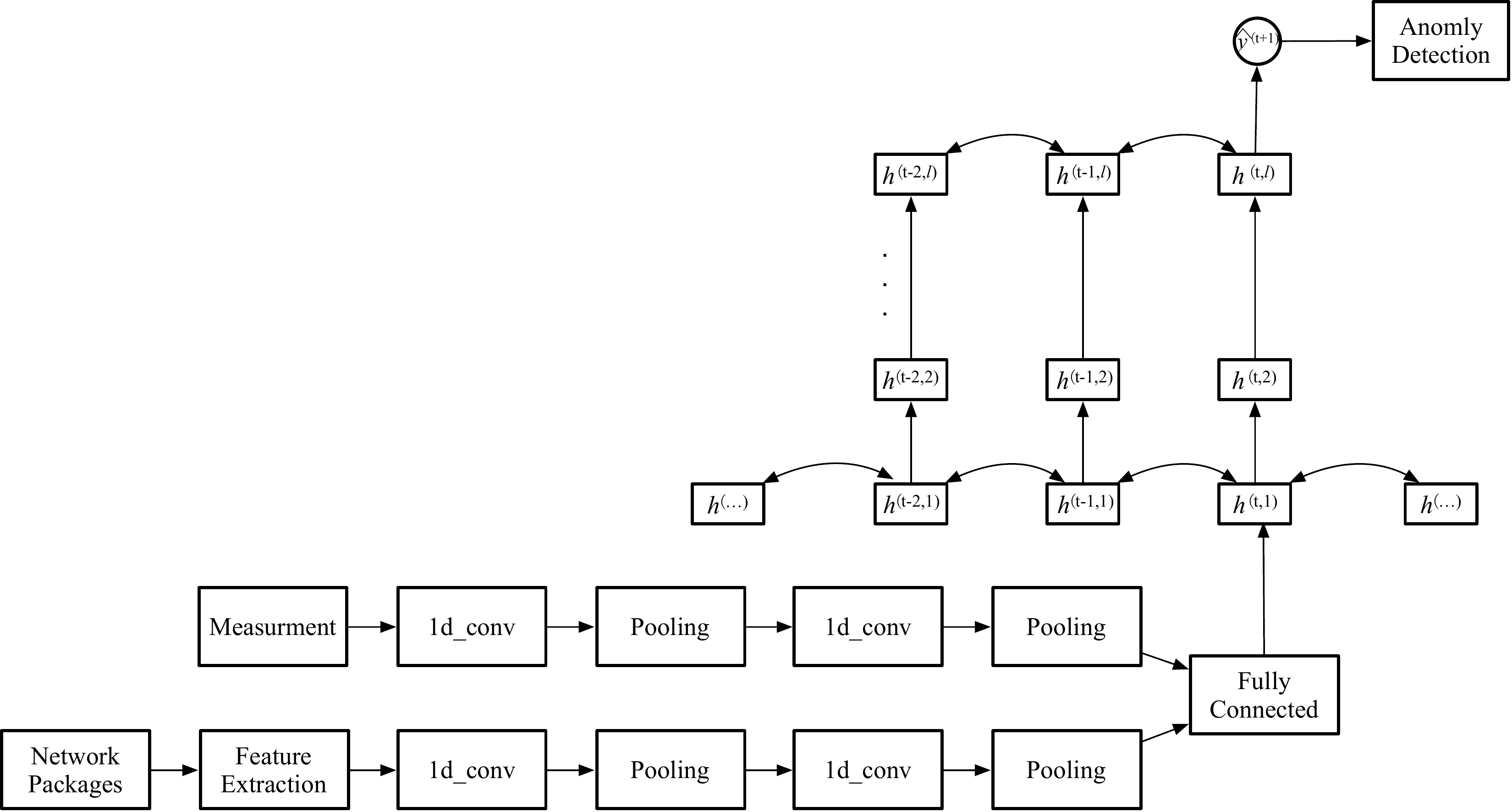}
	\caption{The combined detection method with both network package and data measurement inputs.}
	\label{fig:combined_detection}
\end{figure}

Specifically, the schematic structure of the combined framework is given in Figure \ref{fig:combined_detection}. As seen in the figure, the combined framework is rather straightforward. An alternative method to combine data level information and packet level features is directly concatenate the input vector. However, because the dimension between the data measurements and network packet features differ significantly, direct concatenation may have minimal improvement than aforementioned time-serious methods. Alternatively, each level features are transformed by a convolutional neural network before concatenation as shown in the figure. The purpose of adding additional convolutional neural network is to equalize the dimension between data measurement and packet level features and their respect weights are learned using gradient descent (Adam algorithm is used in our experiments). Inception deep learning architecture \cite{szegedy2015going} is advised when possible.

\section{Case study on IEEE 39 bus system} \label{sec:eva}
\begin{figure}[th]
	\centering
	\includegraphics[width=0.75\linewidth]{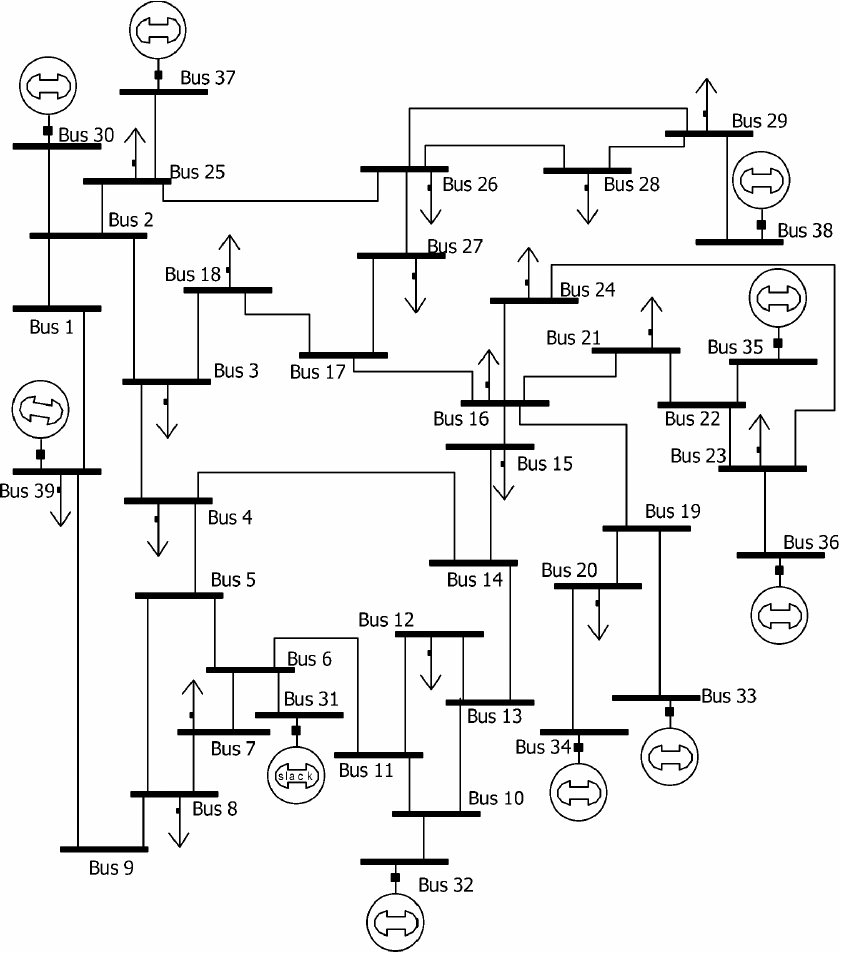}
	\caption{IEEE 39 bus power system.}
	\label{fig:ieee39}
\end{figure}

In this section, we provide several key implement details of our proposed FDI attack detection system, thereby providing a better intuition about its capabilities and limitations. Figure~\ref{fig:ieee39} shows IEEE 10 generator 39 bus power system and details in \cite{pai1989ieee}. In the 39 bus system, the state vector $x \in \mathbb{R}^{39}$ is composed of the voltage, current and frequency of the individual buses. The communication network is emulated using two computers where one computer represents the Independent Service Operator which collect data measurement through Ethernet. The sample rate is set to 10Hz. The FDI attacks are generated from man-in-middle attackers from a client-server communication structure and two input sources are time synchronized to make it possible for real time implementation. The dynamic detector is configured with 3 layers bi-directional RNN with LSTM cells and trained using Pytorch.
In this experiment, to better evaluate our dynamic detector, our system does not implement SE. 

We assume that the attackers can inject $k$ measurements which are randomly chosen to generate Gaussian distributed attack vectors $a \sim \mathcal{N}(0, 0.5)$. We also test the scenario that the attacking vectors are derived from real measurement which will fail to be detected by most state-of-art detectors. In this experiment, the attackers try to inject a false generator trip event which is collected in advance and we define attacking capability as $\frac{k}{n}$ where $n$ is the total number of measurements.
We evaluate the performance of our dynamic FDI attack detection framework on the classification results for the test set. We train a neural network with 10 training epochs to minimize the loss function in Equation \ref{equ:loss}. For the experiment, we apply a 60\% / 20\% / 20\%  train / validation / test split, with a grid search to determine the best $\tau$. 

% \begin{table}[!t]
% \centering
% \caption{Average prediction error for regression and classification datasets.}
% \label{tab_result}
% \begin{tabular}{|c|c|c|c|c|}
% \hline
%          & k=10 & k=20 & k=30 & k=39 \\ \hline
% Two Hidden Layers   & xxx    & xxx     & xxx        & xxx    \\ \hline
% Three Hidden Layers & xxx    & xxx     & xxx        & xxx    \\ \hline
% \end{tabular}
% \end{table}

\begin{figure}[!t]
    \centering
    \includegraphics[width=0.85\linewidth]{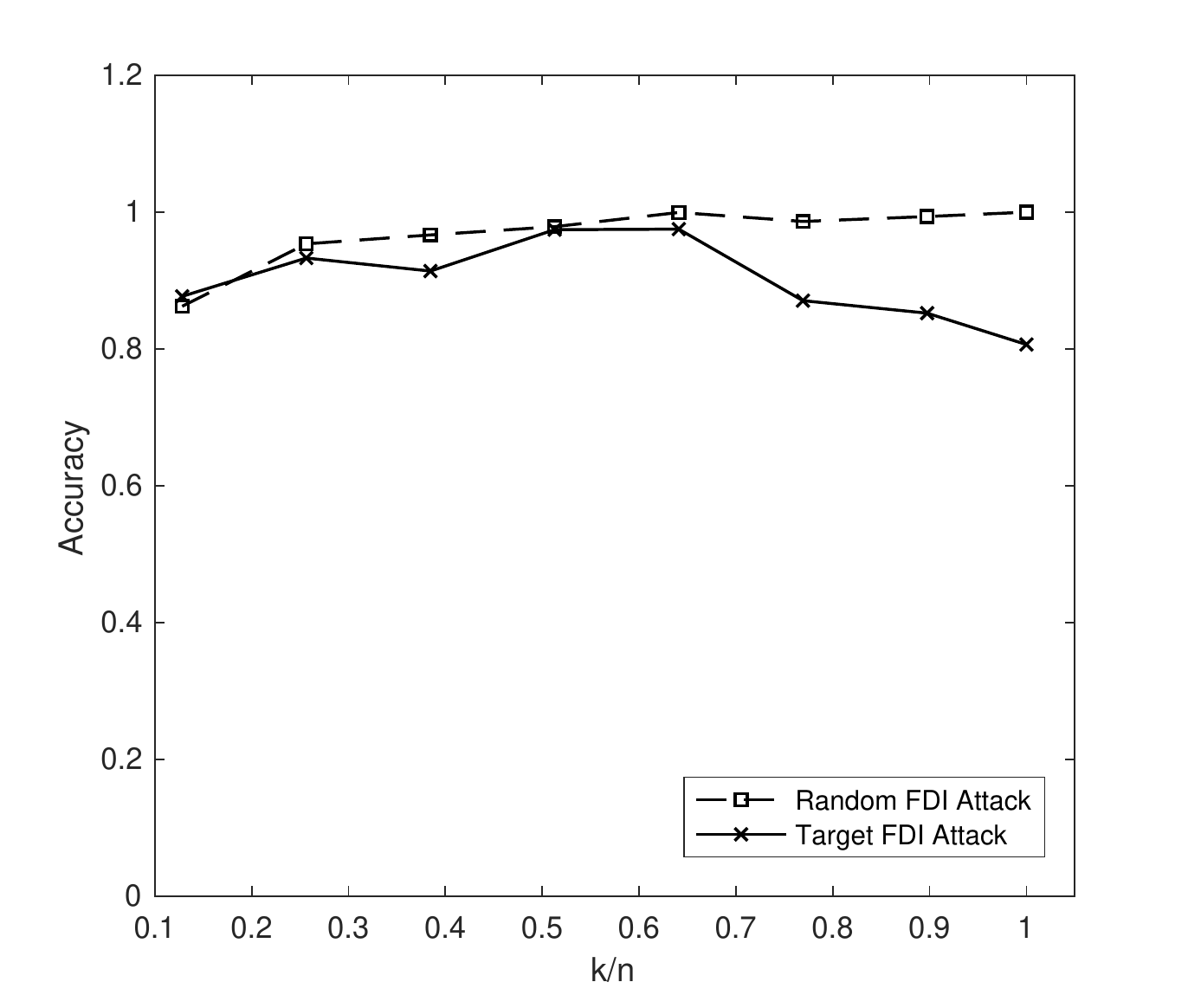}
    \caption{The accuracy of detecting FDI attacks for different the number of the compromised buses.}
	\label{fig:result}
\end{figure}

We illustrate the results of our anomaly detection system in Figure \ref{fig:result}.
From the figure, it is clear that our proposed detection mechanism can achieve the detection accuracy above 90\% for random FDI attacks when $\frac{k}{n}$ is high. However, we also notice that our system has low accuracy when attacking power is low. In fact, this can be resolved by incorporate a SE detector (such as \cite{ozay2016machine,he2017real}) which work well for limited attacking capability. In other words, our proposed two-level detection scheme is able to achieve high detection accuracy for different scenarios. For target FDI attacks, the injected data streams are carefully manipulated from real event which is not considered for most SE bad data detection schemes. Our experiment validates that dynamic features and network anomaly detector integration can support IDS for better performance.
The simulation result in this case study also implies that the full deep knowledge of the power system is not required for the success of our dynamic detection scheme. Our system can be built at early stage of an electricity network.

\section{Conclusion} \label{sec:con}
In this paper, we propose a deep learning based framework to detect measurement anomalies due to FDI attacks. We described our detection methodology that leverages both convolutional neural network and recurrent neural networks. Our model learns normal behavior from normal data and is unrelated to certain attack, and thus can detect unseen attacks. Additionally, our two-level detector is robust using hybrid features and can detect attack when state vector estimator fails.
We provided key insights about various factors that impact the performance of the proposed algorithm. We presented a detailed case study of the proposed algorithm on the IEEE 39-bus system.

\bibliographystyle{plain}
\bibliography{IEEEfull}

\begin{thebibliography}{10}

\bibitem{iec61850}
IEC 61850-90-5.
\newblock Communication networks and systems for power utility automation part
  90-5: use of iec 61850 to transmit synchrophasor information according to
  ieee c37.118, 2012.

\bibitem{ashok2016online}
Aditya Ashok, Manimaran Govindarasu, and Venkataramana Ajjarapu.
\newblock Online detection of stealthy false data injection attacks in power
  system state estimation.
\newblock {\em IEEE Transactions on Smart Grid}, 2016.

\bibitem{bi2014graphical}
Suzhi Bi and Ying~Jun Zhang.
\newblock Graphical methods for defense against false-data injection attacks on
  power system state estimation.
\newblock {\em IEEE Transactions on Smart Grid}, 5(3):1216--1227, 2014.

\bibitem{bobba2010detecting}
Rakesh~B Bobba, Katherine~M Rogers, Qiyan Wang, Himanshu Khurana, Klara
  Nahrstedt, and Thomas~J Overbye.
\newblock Detecting false data injection attacks on dc state estimation.
\newblock In {\em Preprints of the First Workshop on Secure Control Systems,
  CPSWEEK}, volume 2010, 2010.

\bibitem{esmalifalak2014detecting}
Mohammad Esmalifalak, Lanchao Liu, Nam Nguyen, Rong Zheng, and Zhu Han.
\newblock Detecting stealthy false data injection using machine learning in
  smart grid.
\newblock {\em IEEE Systems Journal}, 2014.

\bibitem{feng2017multi}
Cheng Feng, Tingting Li, and Deeph Chana.
\newblock Multi-level anomaly detection in industrial control systems via
  package signatures and lstm networks.
\newblock In {\em Dependable Systems and Networks (DSN), 2017 47th Annual
  IEEE/IFIP International Conference on}, pages 261--272. IEEE, 2017.

\bibitem{fouda2011lightweight}
Mostafa~M Fouda, Zubair~Md Fadlullah, Nei Kato, Rongxing Lu, and Xuemin~Sherman
  Shen.
\newblock A lightweight message authentication scheme for smart grid
  communications.
\newblock {\em IEEE Transactions on Smart Grid}, 2(4):675--685, 2011.

\bibitem{grid2010introduction}
NIST~Smart Grid.
\newblock Introduction to nistir 7628 guidelines for smart grid cyber security.
\newblock {\em Guideline, Sep}, 2010.

\bibitem{hao2016likelihood}
Yingshuai Hao, Meng Wang, and Joe~H Chow.
\newblock Likelihood analysis of cyber data attacks to power systems with
  markov decision processes.
\newblock {\em IEEE Transactions on Smart Grid}, 2016.

\bibitem{he2017real}
Youbiao He, Gihan~J Mendis, and Jin Wei.
\newblock Real-time detection of false data injection attacks in smart grid: A
  deep learning-based intelligent mechanism.
\newblock {\em IEEE Transactions on Smart Grid}, 8(5):2505--2516, 2017.

\bibitem{hochreiter1997long}
Sepp Hochreiter and J{\"u}rgen Schmidhuber.
\newblock Long short-term memory.
\newblock {\em Neural computation}, 9(8):1735--1780, 1997.

\bibitem{kim2011strategic}
Tung~T Kim and H~Vincent Poor.
\newblock Strategic protection against data injection attacks on power grids.
\newblock {\em IEEE Transactions on Smart Grid}, 2(2):326--333, 2011.

\bibitem{krizhevsky2012imagenet}
Alex Krizhevsky, Ilya Sutskever, and Geoffrey~E Hinton.
\newblock Imagenet classification with deep convolutional neural networks.
\newblock In {\em Advances in neural information processing systems}, pages
  1097--1105, 2012.

\bibitem{liang2017review}
Gaoqi Liang, Junhua Zhao, Fengji Luo, Steven~R Weller, and Zhao~Yang Dong.
\newblock A review of false data injection attacks against modern power
  systems.
\newblock {\em IEEE Transactions on Smart Grid}, 8(4):1630--1638, 2017.

\bibitem{liu2014detecting}
Lanchao Liu, Mohammad Esmalifalak, Qifeng Ding, Valentine~A Emesih, and Zhu
  Han.
\newblock Detecting false data injection attacks on power grid by sparse
  optimization.
\newblock {\em IEEE Transactions on Smart Grid}, 5(2):612--621, 2014.

\bibitem{liu2011false}
Yao Liu, Peng Ning, and Michael~K Reiter.
\newblock False data injection attacks against state estimation in electric
  power grids.
\newblock {\em ACM Transactions on Information and System Security (TISSEC)},
  14(1):13, 2011.

\bibitem{ozay2016machine}
Mete Ozay, Inaki Esnaola, Fatos Tunay~Yarman Vural, Sanjeev~R Kulkarni, and
  H~Vincent Poor.
\newblock Machine learning methods for attack detection in the smart grid.
\newblock {\em IEEE Transactions on Neural Networks and Learning Systems},
  27(8):1773--1786, 2016.

\bibitem{pai1989ieee}
MA~Pai, T~Athay, R~Podmore, and S~Virmani.
\newblock Ieee 39-bus system.
\newblock 1989.

\bibitem{rumelhart1986learning}
David~E Rumelhart, Geoffrey~E Hinton, and Ronald~J Williams.
\newblock Learning representations by back-propagating errors.
\newblock {\em nature}, 323(6088):533, 1986.

\bibitem{schuster1997bidirectional}
Mike Schuster and Kuldip~K Paliwal.
\newblock Bidirectional recurrent neural networks.
\newblock {\em IEEE Transactions on Signal Processing}, 45(11):2673--2681,
  1997.

\bibitem{szegedy2015going}
Christian Szegedy, Wei Liu, Yangqing Jia, Pierre Sermanet, Scott Reed, Dragomir
  Anguelov, Dumitru Erhan, Vincent Vanhoucke, Andrew Rabinovich, et~al.
\newblock Going deeper with convolutions.
\newblock Cvpr, 2015.

\bibitem{tavallaee2009detailed}
Mahbod Tavallaee, Ebrahim Bagheri, Wei Lu, and Ali~A Ghorbani.
\newblock A detailed analysis of the kdd cup 99 data set.
\newblock In {\em Computational Intelligence for Security and Defense
  Applications, 2009. CISDA 2009. IEEE Symposium on}, pages 1--6. IEEE, 2009.

\end{thebibliography}

\end{document}